\documentclass[12pt,preprintnumbers,aps,amssymb,nofootinbib]{revtex4}
\usepackage{epsfig}
\usepackage{graphicx}
\usepackage{graphicx}
\usepackage{epstopdf}
\def\eq#1{{(\ref{#1})}}
\def\fig#1{{Fig.~\ref{#1}}}

\newcommand{\beq}{\begin{equation}}
\newcommand{\eeq}{\end{equation}}
\newcommand{\beqar}[1]{\begin{eqnarray}\label{#1}}
\newcommand{\eeqar}{\end{eqnarray}}

\newcommand{\as}{\alpha_s}
\newcommand{\un}{\underline}

\newcommand{\jpsi}{{J/\Psi}}
\newcommand{\lqcd}{\Lambda_\mathrm{QCD}}
\newcommand{\stackeven}[2]{{{}_{\displaystyle{#1}}\atop\displaystyle{#2}}}
\newcommand{\lsim}{\stackeven{<}{\sim}}
\newcommand{\gsim}{\stackeven{>}{\sim}}

\begin{document}

\preprint{BNL-NT/05-41}

\title {
Signatures of the Color Glass Condensate \\ in $J/\Psi$ production 
off nuclear targets
} 


\author{Dmitri Kharzeev}

\affiliation{ Physics Department, Brookhaven National Laboratory,\\
Upton, NY 11973-5000, USA}

\author{Kirill Tuchin}
 
\affiliation{ Physics Department and RIKEN-BNL Research Center, Brookhaven National Laboratory,\\
Upton, NY 11973-5000, USA\\ and \\
Department of Physics and Astronomy, Iowa State University,
Ames, IA, 50011
}

\date{\today}

\begin{abstract} 
We consider the $J/\Psi$ production in proton (deuteron)
-- nucleus collisions at high energies.  We argue that the production
mechanism in this case is different from that in $pp$ collisions due to gluon 
saturation in the nucleus and formation of the Color Glass Condensate.  
At forward rapidities (in the proton fragmentation region), the production 
of $J/\Psi$ is increasingly suppressed both as a function of rapidity and
centrality. On the other hand, at backward rapidities at RHIC (in the
fragmentation region of the nucleus) the coherent effects lead to a modest 
enhancement of the production cross section, with the nuclear modification
factor $R_{J/\Psi}$ increasing with centrality.
We find that the $J/\Psi$ production cross section exhibits at forward rapidities the
limiting fragmentation scaling established previously for soft processes; in the energy range 
studied experimentally, it manifests itself as an approximate ``$x_F$ scaling". 

\end{abstract}
\maketitle 


\section{Introduction}

Understanding the $\jpsi$ production mechanism is one of the challenges of
QCD. On one hand the charm quark mass is quite large on the typical QCD scale of $\lqcd$, which
makes the use of perturbative QCD meaningful \cite{collcharm} since the long
distance dynamics is effectively decoupled \cite{Appelquist:tg}.  However
the size of this system and the inverse of the binding energy are not
small enough to suppress significant non-perturbative contributions.  
Indeed, perturbative QCD fails in describing the differential $\jpsi$
production cross section and the polarization. Different mechanisms were
suggested to explain the existing experimental data. Unfortunately all of
them so far have encountered problems in describing at least some of the observables 
(for a recent review, see \cite{Brambilla:2004wf}).
In the context of high energy nuclear physics, it is important to understand well the mechanism of $\jpsi$
production also in nuclear processes since $\jpsi$ suppression in heavy-ion
collisions could serve as a signal of the Quark-Gluon Plasma formation
\cite{Matsui:1986dk}.

One of the long-standing puzzles is the lack of $x_2$ scaling of the 
nuclear modification factor ($x_2$ is
the Bjorken variable corresponding to the nuclear target parton distribution)
in $J/\psi$ production off nuclei.  Even though this scaling is
expected to hold in the parton model, the data from CERN \cite{Badier:1983dg} and FNAL \cite{Leitch:1999ea} fixed
target experiments are in violent contradiction with this expectation. 
The absence of $x_2$ scaling has become even more dramatic at RHIC \cite{Adler:2005ph}.
Instead of the badly broken $x_2$ scaling, the data instead exhibit an
unexpected approximate scaling in the Feynman $x_F$ variable.

It was realized long time ago \cite{Brodsky:1991dj} that this lack of $x_2$ scaling, and thus the violation of 
QCD factorization, is caused by multi-parton (higher twist) interactions in the nuclear target. Several specific mechanisms 
of this type were considered over the years \cite{Kopeliovich:1984bf,Vogt:1991qd,Gavin:1991qk,Kharzeev:1993qd,Clavelli:1985kg,Benesh:1994du,Fujii:2003ff}.

In this paper we would like to re-visit the problem of $\jpsi$ production
in proton -- nucleus collisions at high energy basing on the novel Color Glass
Condensate picture of the nuclear wave function at small $x$. In this approach, the 
strength of the color field inside the nucleus is proportional to the saturation scale $Q_s^2(x_2)$
determined by the density of partons in the transverse plane.
It is a growing function of the collision energy and the atomic number of the nucleus.
Experimental data indicate that at RHIC kinematics $Q_s\gg \lqcd$ which
implies that the inter-nucleon interactions play a little role in pA 
interactions at high energies. Therefore, at high energies a nuclear color 
field can be described by only one universal (process independent) 
dimensional scale $Q_s(x_2)$ \cite{GLR,Mueller:wy,Blaizot:nc,MV}.
The production of heavy quarks in this framework has been previously considered 
in several papers \cite{KhT,KTcc,Gelis:2003vh,Blaizot:2004wv,Gelis:2004jp,Fujii:2005vj}.

In the previous publications \cite{KhT,KTcc}, we have argued that there are
two different dynamical regimes of heavy quark production at high energies
depending on the relation between the saturation scale and the quark's
mass $m$. When $Q_s\ll m$ the heavy quark production is incoherent,
meaning that it is produced in a single sub-collision of a proton with a 
nucleon. This case can be treated within a conventional
perturbative approach. In the opposite limit of $Q_s\gg m$ the heavy quark
production is coherent since the whole nucleus takes part in the process.
In this case the heavy quark production is sensitive to
a strong color field (CGC) which violates the decoupling
of the subprocess of heavy quark production from the dynamics of partons in 
the
nuclear wave function \cite{LRSS,KhT}. The reason is that the decoupling
theorems can be applied only when the heavy quark mass is much larger than
the typical hadronic scale, which is of the order of $Q_s(x)$ at high
energies.

The goal of this paper is to address the problem of $\jpsi$ production at
high energies. It is organized as follows. In section \ref{sec:coh} we 
argue that at high energies the
time of interaction of the projectile proton with the target nucleus at
rest is much smaller than the time of heavy quark pair production and
subsequent formation of a bound state. This will allow us to use the
eikonal approximation and derive in section \ref{sec:xsec} the cross
section for the $\jpsi$ production in pA collisions at high energies. In
section \ref{sec:interplay} we study the derived expression in two
different kinematical regions. We show that at $Q_s\ll M_\psi$ the $\jpsi$ the 
cross section is an increasing function of centrality since the dominant
contribution comes from the two-gluon exchange process. At $Q_s\gg M_\psi$
multiple re-scattering as well as quantum evolution lead to the suppression of
$\jpsi$ production both as a function of energy/rapidity and centrality.  
We also point out that the approximate $x_F$ scaling observed in SPS and FNAL data 
emerges naturally in our approach. However, it is seen to be a consequence of the 
slow dependence of the gluon distribution on energy, and so is broken 
at higher energies. However, as the energy increases (e.g. to the LHC energy range) the $x_F$ scaling 
is restored owing to the onset of gluon saturation in the incident {\it proton}.
We compare our calculations with the available experimental data in Sec.~ \ref{sec:compar}
and conclude in Sec.~\ref{sec:summary}.

\section{Different production mechanisms of $\jpsi$ on nucleus}\label{sec:coh}
\subsection{Relevant time scales}

Consider charmonium production in a pA collision. A $c\bar c$ pair is 
produced over the time $1/(2m_c)$ in its center-of-mass frame. In the nucleus  
rest frame this time is Lorentz-dilated \cite{MuBr,KZ}
\beq\label{prod1}
\tau_P\,\approx\, \frac{1}{(2\, m_c)}\,\frac{E_g}{(2\, m_c)}\,,
\eeq
where $E_g$ is a parent gluon's energy.
This time scale should be compared to the typical interaction time 
$\tau_\mathrm{int}\sim R_A/c$. Eq.~\eq{prod1} shows that at very high 
energies the production time of $c\bar c$ pair can be much larger 
than the interaction time $\tau_P\,\gg\,\tau_\mathrm{int}$. This a general 
property of all hard processes 
at high energies: they develop over a long time $\tau_P$ \footnote{ 	
Sometimes one introduces the ``coherence length" $l_c=\tau_P/c$.} 
\cite{Ioffe}.

This formula can be rewritten in terms of the Bjorken variable 
associated with nucleus, $x_2$. Note, that the gluon takes fraction $x_1$ 
of the proton's energy $E_g\,=\,x_1\, E_p$. Also, by four-momentum 
conservation, $(2\, m_c)^2\,=\,x_1\, x_2\, s\,=\,2\,x_1\, x_2\, M_N\, 
E_p$, where $M_N$ is the nucleon mass and $E_p$ is the proton's energy. 
Thus, it follows from \eq{prod1} that 
\beq\label{prod2}
\tau_P\,\approx\, \frac{1}{2\, M_N\, x_2}\,.
\eeq
At RHIC, in the center-of-mass frame $x_2\,=\, 
(m_c/\sqrt{s})\,e^{-y}\,=\,6.5\cdot 10^{-3}\,e^{-y}$, where we 
introduced rapidity $y$. 
Therefore,
\beq\label{prodRHIC}
\tau_P(\mathrm{RHIC})\,\approx\,15\,e^y\,\mathrm{fm}.
\eeq
Equation \eq{prodRHIC} implies that at forward rapidities $y>1$ one can 
indeed assume that the proton interacts coherently with the whole nucleus (similar estimates 
for the fixed target energies can be found in \cite{Kharzeev:1995id}).
In this case the transverse size of the $c\bar c$ pair  
is fixed during its propagation through the nucleus and we can apply the 
eikonal approximation for calculation of the scattering amplitude 
\cite{NZ,AMdipole}.  

Finally, the $\jpsi$ wave function is formed from the initial $c\bar c$ pair. This lasts 
$\sim\,2/(M_{\Psi'}-M_\Psi)$ in the $\jpsi$ rest frame. In the nucleus 
rest frame  the $\jpsi$ formation time is \cite{MuBr,KZ}
\beq\label{form}
\tau_F\,\approx\, \frac{2}{M_{\Psi'}-M_\psi}\,\frac{E_g}{M_\psi}\,.
\eeq
Since the $\jpsi$ bounding energy is much less than its mass, the $\jpsi$ 
production time is much shorter than its formation time: $\tau_P\,\ll\, 
\tau_F$. A more accurate evaluation of the formation time can be performed with the help 
of the spectral representation for the $\jpsi$ propagator by using the experimental data on 
$e^+ e^-$ annihilation into charm quarks \cite{Kharzeev:1999bh}; this leads to the $\jpsi$ proper 
formation time of $0.45$~fm.
Thus, at RHIC 
\beq\label{formRHIC}
\tau_F(\mathrm{RHIC})\,\approx\, 41\, e^y\,\mathrm{fm};
\eeq  
the $\jpsi$ wave function is therefore formed outside of the nucleus at rapidities 
$y\,\gsim\,-2$. 

The relationships between the three relevant time scales $\tau_\mathrm{int}$,
$\tau_P$ and $\tau_F$ depend on the collision energy $\sqrt{s}$ and the 
rapidity $y$. In the next subsection we classify all possible situations 
in the RHIC kinematical region ($\sqrt{s}\,=\,200$~GeV).

\subsection{Coherent versus incoherent $\jpsi$ production}\label{sec:vs}
\subsubsection{Forward rapidities}

\begin{figure}
\begin{center}
\epsfig{file=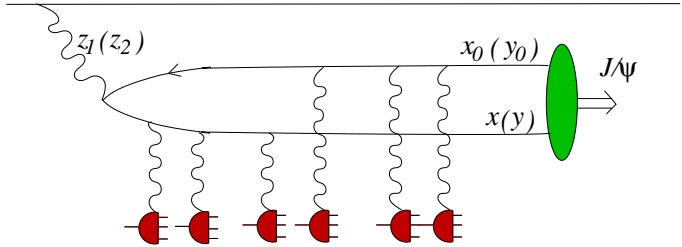,width=9cm}
\caption{Production of $\jpsi$ in pA collisions at high energy. At RHIC
$\sqrt{s}=200$~GeV: $y\gsim 2$.} \label{jpsiR}
\end{center}
\end{figure}

At (pseudo)-rapidities $y\,\gsim\,1$ 
$\tau_F\,\gg\,\tau_P\,\gg\tau_\mathrm{int}$. This implies that the $c\bar 
c$ pair scatters coherently off all the nucleons along its trajectory in the 
nucleus, see \fig{jpsiR}. Therefore, the 
process of $\jpsi$ formation proceeds through the following three stages in the nucleus rest frame. 
First, way before the collision with the 
nucleus, the fast proton develops a cloud of virtual partons which includes 
one $c\bar c$ pair (in the leading order in $\as$). In the light-cone 
perturbation theory this is described by the valence quark and the virtual 
gluon wave functions which are explicitly displayed in the next section, 
see \ref{subsec:wf}.  
Second, the coherence of the cloud is destroyed by interaction  
with nucleons in nucleus. Due to the large production time $\tau_P$  
the scattering matrix can be diagonalized in the color 
dipole basis: the transverse size  of the $c\bar c$ is 
fixed during the interaction. We calculate the $c\bar cA$ amplitude in section 
\ref{subsec:prop}. Third, $\jpsi$ is formed far away from the nuclear 
remnants. No nuclear effects are expected at this stage. 

\subsubsection{Central rapidities}

\begin{figure}
\begin{tabular}{ccc}
\epsfig{file=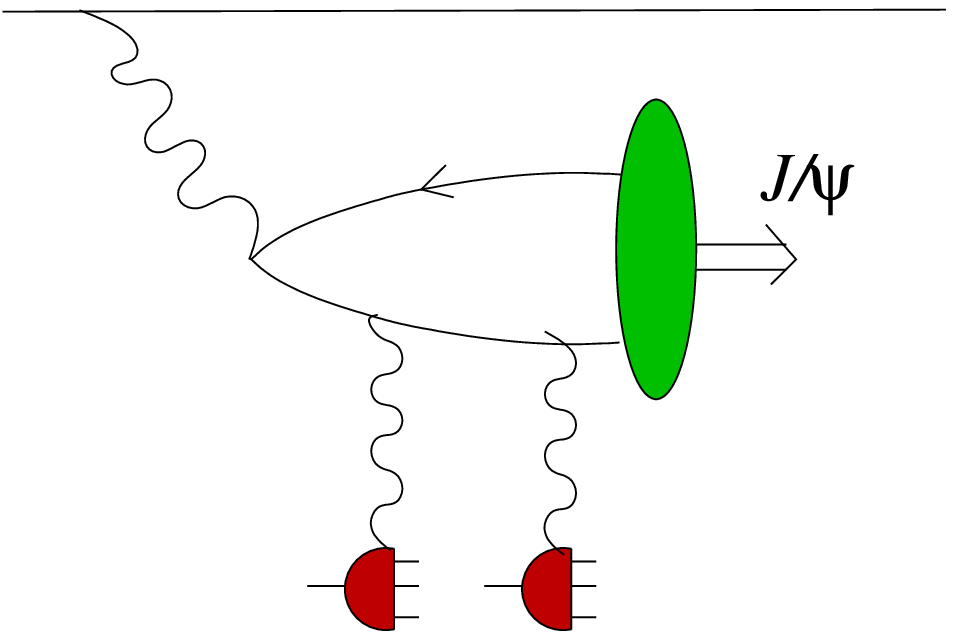,width=5cm}& \mbox{}\hspace{1cm}&
\epsfig{file=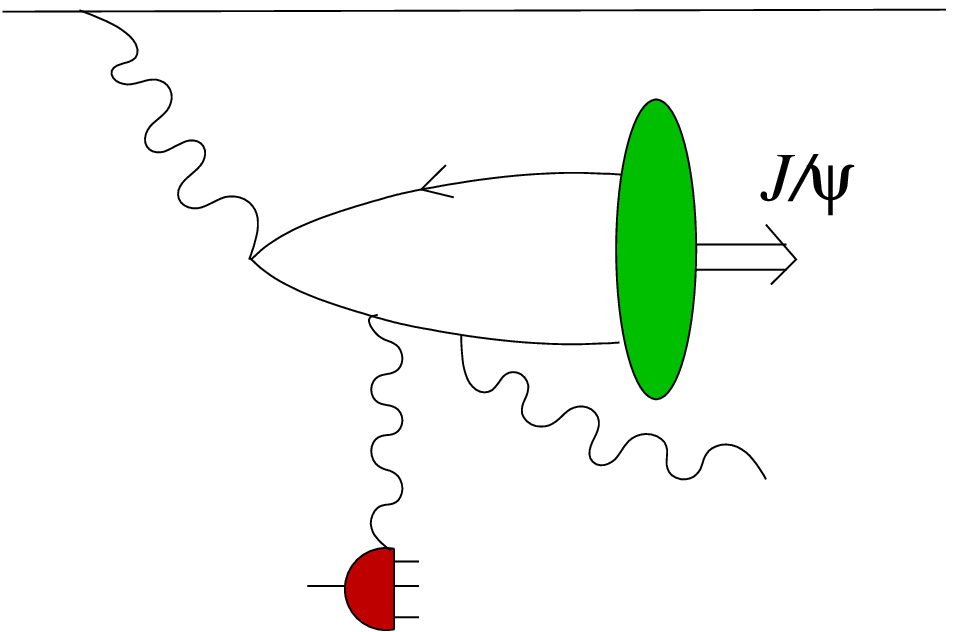,width=5cm}\\
(a)&&(b)
\end{tabular}
\caption{Two mechanisms of $\jpsi$ production:  (a) by two-gluon exchange;
(b) by one-gluon exchange and one-gluon emission.}
\label{prod_jpsi}
\end{figure}

At rapidities $-1\,\lsim\,y\,\lsim\,0$ the  production time becomes 
smaller than the nuclear size which implies that the $c\bar c$ pair 
scatters coherently off a few nucleons. This leads to the \emph{enhancement} of 
$\jpsi$ production since in that case the main contribution 
to the scattering amplitude arises from the diagram shown in 
\fig{prod_jpsi}(a) which is enhanced by an additional power of $A^{1/3}$ 
with respect to the diagram \fig{prod_jpsi}(b) which describes $\jpsi$ 
production in pp collisions.
 
The diagram \fig{prod_jpsi}(a), where the $\jpsi$ is produced by double 
gluon exchange is parametrically enhanced compared to the diagram shown in 
\fig{prod_jpsi}(b) where the $\jpsi$ is produced by one gluon exchange. 
Indeed, let us for a moment concentrate on $\jpsi$ production in a 
quasi-classical
approximation where the coupling is small $\as(Q_s^2)\ll 1$ and together
with atomic number $A\gg 1$ it forms a resummation parameter
$\as^2 A^{1/3}\sim 1$ \cite{YuK}. The diagram (b) in \fig{prod_jpsi}
is of the order $\as^5 A^{1/3}\sim \as^3$ while the diagram (a) is of
the order of $\as^6 A^{2/3}\sim \as^2$. Therefore, the diagram (a) is
enhanced provided that the nucleus is sufficiently large.

This conclusion remains valid beyond the quasi-classical approximation. 
The gluons emitted in the course of quantum evolution get resummed into gluon 
distribution functions, which therefore grow fast as $x$ decreases. The 
resummation parameter $\as\, xG(x,Q^2)\, A^{1/3}$ becomes large even for the  
proton \cite{GLR}. At small enough 
$x$ the diagram (a) dominates the $\jpsi$ production even for the scattering 
off light nuclei $A\sim 1$. The general effect of the enhancement of double-gluon 
exchange diagrams in hard processes on heavy nuclei has been already pointed out 
by  in \cite{Brodsky:1991dj,HKZ}. 

Although the dipole model qualitatively describes the effect of the 
enhancement of $\jpsi$ production at $-1\,\lsim\,y\,\lsim \,0$  it cannot 
be applied at $y\,\lsim\,0$ due to the effect of finite production time.
In other words to get a reasonable description of the process  
one has to include the absorption corrections of the color dipole $c\bar 
c$ in the nuclear medium (note that $\tau_F$ is still much larger than 
$R_A$). In the present publication we are going to analyze only $y>0$ 
region at RHIC. However when comparing to the experimental data from CERN and 
FNAL we will have to correct the results by the absorption factors, see 
Sec.~\ref{sec:att}.

To quantify the effect of the finite production time and to specify 
 the region of 
applicability of eikonal approximation, we can consider the longitudinal 
nuclear form factor $F_A^2(q_z)$, 
which takes into account the quantum interference in the longitudinal direction 
at a finite longitudinal momentum transfer \cite{Gribov:1969zy} (this formfactor is assumed to equal unity   
in the dipole model). It is defined as 
\beq\label{ff}
F_A^2(q_z)\,=\,A^{-1}\,\int d^2b\, \left| \,\int_{-\infty}^\infty dz\, 
\rho(b,z)\,
e^{i\,q_z\,z}\,\right|^2\,,
\eeq
where $q_z$ is the longitudinal momentum transfer and 
$\rho(b,z)$ is the density of nucleons in a nucleus. The typical value of 
the longitudinal coordinate in \eq{ff} is $z\sim R_A$. Therefore, the 
integral over $z$ vanishes due to rapid oscillations of the exponential 
factor unless $q_z\,< \, 1/R_A$. On the other hand $q_z\,\sim\,1/\tau_P$. 
Thus, $F_A(q_z)$ is sizable  (it is normalized to unity) when 
$\tau_P\,>\,R_A$ and is small otherwise. 

\begin{figure}
\epsfig{file=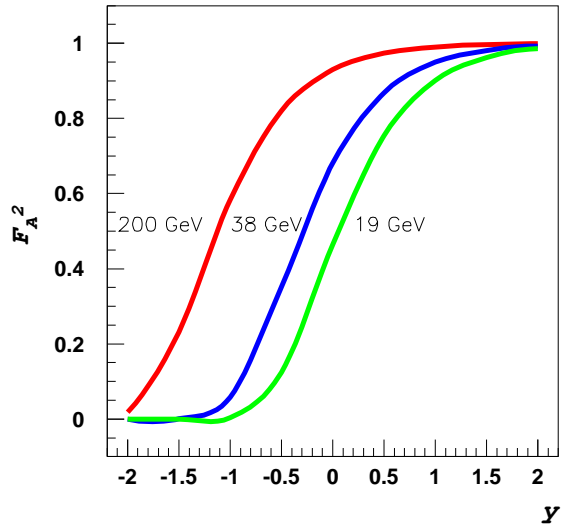,width=9cm}
\caption{The longitudinal form factor $F_A^2$ as a function of rapidity 
$y$ for RHIC, Tevatron and SPS energies.}
\label{abc}
\end{figure}

Assuming for simplicity a 
Gaussian parameterization of the nuclear density we obtain $F_A^2(q_z)\,=\,
e^{-R_A^2\, q_z^2/3}$ \cite{HKZ}. In the \fig{abc} we plot the 
longitudinal form factor for different energies.
We learn from \fig{abc} that the corrections to the dipole model due to a 
finite $\tau_P$ are of the order of 10\% at $y=0$ at RHIC 
($\sqrt{s}=200$~GeV), but already at $y=-1$ they are as large as 50\%. 
Therefore, the dipole model should provide an accurate result for 
$y\,\gsim\, 0$ which correspond to $x_F\,\gsim\, 0$. At
Fermilab fixed target experiments ($\sqrt{s}=38$~GeV) the dipole 
model is applicable (corrections $\lsim$ 10\%) at rapidities larger than 
$\sim \ln(200/38)= 1.7$ 
which correspond to $x_F\,\simeq\,0.4$. At SPS ($\sqrt{s}\,=\,19$~GeV) the dipole model 
has to be corrected even in the most forward region.

\subsubsection{Backward rapidities}

At rapidities $y\,\lsim\,- 2$ the coherence is completely lost and the 
process of $\jpsi$ production in pA collisions becomes similar to the one 
in pp collisions. All of the dependence on $A$ arises then from the propagation of the produced $c\bar 
c$ pair and $\jpsi$ through the nuclear matter.

\section{Cross section for $\jpsi$ production at forward 
rapidities.}\label{sec:xsec}

\subsection{Propagation of the $c\bar  c$ pair through the 
nucleus}\label{subsec:prop}

The first step in calculation of the $\jpsi$ production cross section, see 
\fig{jpsiR} , 
is the calculation of the scattering amplitude of $q_vc\bar c$ state off the 
nucleus with the projection on the color neutral state after the $c\bar c$ 
passes the nucleus. In our analysis we will assume, for the sake of 
simplicity, that the valence quark is a spectator. 

It is straightforward, although laborious, to calculate all possible gluon 
attachments to quark and antiquark in \fig{jpsiR}. 
In the large $N_c$ approximation 
the $c\bar c$ amplitude for a single scattering (\fig{prod_jpsi}) 
projected on the color singlet state reads  
\beq\label{oneresc}
\mathcal{M}_1(\un x, \un x_0, \un y, \un y_0)=
\frac{1}{4}\bigg(\frac{Q_s^2}{4}\bigg)^2\, \bigg[(\un x-\un y)^2\,+\,(\un 
x_0-\un y_0)^2\bigg]^2\,.
\eeq 

The saturation scale $Q_s^2$ in \eq{xi} is given by \cite{AlMu}
\beq\label{satscale}
Q_s^2(\un x)=\frac{4\pi^2\as N_c}{N_c^2-1}\,\rho\, T(\un b)\,
xG(x,1/\un x^2),
\eeq 
where the gluon distribution function in the nucleon reads
\beq\label{xG}
xG(x,1/\un x^2)=\frac{\as\, C_F}{\pi}\,\ln\frac{1}{\un x^2\mu^2},
\eeq
where $\mu$ is some infrared cutoff. For a spherical sharp-edge nucleus $T(\un b)=
2\sqrt{R^2-\un b^2}$.

Assuming that the scatterings off the individual nucleons are independent, one can 
 generalize \eq{oneresc} to include multiple rescatterings (see 
\fig{jpsiR}) in a 
straightforward way. The  $q\bar qA$ amplitude then takes the form 
\beq\label{xi}
\mathcal{M}(\un x, \un x_0, \un y, \un y_0)=1\,-\, \exp\bigg\{
-\frac{1}{4}\bigg(\frac{Q_s^2}{4}\bigg)^2\, \bigg[(\un x-\un 
y)^2\,+\,(\un x_0-\un y_0)^2\bigg]^2
\bigg\}
\eeq
We checked that the color factor exponentiates using the FeynCalc package \cite{feyncalc}.

\subsection{Wave functions}\label{subsec:wf}
In view of the arguments given in the previous section,  at forward rapidities 
the processes of formation of the 
proton wave function, the subsequent rescatterings of the proton's partons inside the  
target nucleus and the formation of the bound $c\bar c$ state are well 
separated in time in the nucleus rest frame. Hence, to proceed we 
need to 
know the light-cone wave functions of the valence quark, the virtual gluon and the 
$\jpsi$. In the $A_+=0$ light cone gauge the valence quark's wave function 
in the configuration space reads \cite{AMdipole}
\beq\label{vquarkc}
\Psi_{q_v\rightarrow q_v g}(\un z)=\int\,\frac{d^2q}{(2\pi)^2}\, 
e^{-i\,\un q \cdot \un z}\, \Psi_{q_v\rightarrow q_v g}(\un q)=
g\, T^a\,\frac{1}{2\pi i}\, \frac{\un\epsilon^\lambda\cdot \un z }{\un
z^2},
\eeq
Averaging the square of Eq.~\eq{vquarkc} over the quantum numbers of the
initial quark and summing over the quantum numbers of the final quark
and gluon we obtain the familiar gluon radiation kernel of a dipole
model
\beq\label{lcv}
\Phi_{q_v\rightarrow q_v g}(\un z_1, \un z_2)=
\frac{1}{2N_c}\,\sum_{a,\lambda}
\Psi_{q_v\rightarrow q_v g}(\un z_1)\Psi^*_{q_v\rightarrow q_v g}(\un 
z_2)=
\frac{\as C_F}{2\pi}\,
\frac{\un z_1\cdot \un z_2}{\un z_1^2\,\un z_2^2},
\eeq
where $\un z_1$ and $\un z_2$ are the transverse coordinates of the
gluon in the amplitude and in the complex conjugated amplitude
correspondingly, see \fig{jpsiR}.

The light-cone wave function of a virtual gluon of momentum $q$ reads 
\cite{NZ,Kopel1,KTcc}
\beq\label{vgluonm}
\Psi_{g^*\rightarrow q\bar q}(\un k, \un k-\un q,\alpha)=
\frac{g\,T^a}{(\un k-\alpha\, \un q)^2+m^2}\,(\delta_{r,r'}(\un k-
\alpha \,\un q)\cdot \un
\epsilon^\lambda\,[r(1-2\alpha)+\lambda]+r\,\delta_{r,-r'}\,m\,(1+r\lambda)),
\eeq
where $\un k$ is the produced quark's transverse momentum, $m$ its mass,
$\alpha=k^+/q^+$ is the
fraction of the gluon's light-cone momentum $q^+$ it carries,
$r$ and $r'$ are the quark and the antiquark helicities
correspondingly. Projecting it onto the vector meson wave function 
\cite{BFGMS} and summing over the polarization and helicity indices one 
can find the overlap function $\Psi_V(\un k,\alpha)\ast \Psi_{g^*}(\un 
k,\alpha)$ in momentum space \cite{RRML} which can be Fourier transformed 
to configuration space. In the non-relativistic 
approximation $\Psi_V(\un r,\alpha)\propto \delta(\un r)\,\delta(\alpha- 
\frac{1}{2})$ the 
overlap function in the configuration space takes the form \cite{GLLMN} 
\beq\label{lcg}
\Psi_\psi\ast \Psi_{g^*}(\un r,\alpha=\frac{1}{2})=
\sqrt{\frac{3\,\Gamma_{ee}\,M_\psi}{48\,\alpha_\mathrm{em}\,\pi}}\,
\frac{m^3\, \un r^2}{4}\, K_2(mr)\,,
\eeq
where $\un r=\un x-\un x_0$, $K_2$ is the modified Bessel function, 
$\Gamma_{ee} \simeq 5.26$~KeV is the leptonic width of $\jpsi$ and we will include 
two delta functions directly into the expression for the cross section.

\subsection{Cross section}

Using \eq{lcv}, \eq{lcg},  and \eq{xi} we can obtain the inclusive $\jpsi$ 
production cross section 
\begin{eqnarray}\label{main_resc}
\frac{d\sigma}{d^2p \,dy}&=&
\int d^2b\, \int\,d^2z_1\, d^2z_2\,\frac{\as\, C_F}{\pi^2}\,
\frac{\un z_1\cdot \un z_2}{\un z_1^2\un z_2^2}\,
\int \frac{d^2x\, d^2 y}{(2\pi)^3}\,
\int \frac{d^2x_0\, d^2 y_0}{(2\pi)^3}
\int d\alpha\int d^2u\, d^2v\,\nonumber\\
&&
\times\Psi_\psi\ast\Psi_{g^*}(\un x-\un x_0,\alpha)\,
\Psi^*_\psi\ast\Psi^*_{g^*}(\un y-\un y_0,\alpha)\,
\,\mathcal{M}(\un x, \un x_0\, \un y, \un y_0)\, 
e^{-i\un p\cdot (\un u-\un v)}\nonumber\\
&&
\times
\delta((\un x_0-\un z_1)+\alpha(\un x-\un x_0))\,
\delta((\un y_0-\un z_2)+\alpha(\un y-\un y_0))\nonumber\\
&&
\times
\delta((\un x_0-\un u)+\alpha(\un x-\un x_0))\,
\delta((\un y_0-\un v)+\alpha(\un y-\un y_0))
\end{eqnarray}
where $\un u$ and $\un v$ are the $\jpsi$ coordinates in the amplitude 
and in the complex-conjugated one respectively. 
Performing the integration 
over  $\un u$ and $\un v$ using the last two delta-functions in 
\eq{main_resc} somewhat simplifies this expression:
\begin{eqnarray}\label{main_resc2}
\frac{d\sigma}{d^2k \,dy}&=&
\int d^2b\, \int\,d^2z_1\, d^2z_2\,\frac{\as\, C_F}{\pi^2}\,
\frac{\un z_1\cdot \un z_2}{\un z_1^2\un z_2^2}\,
\int \frac{d^2x\, d^2 y}{(2\pi)^3}\,
\int \frac{d^2x_0\, d^2 y_0}{(2\pi)^3}
\int d\alpha\nonumber\\
&&
\times\Psi_\psi\ast\Psi_{g^*}(\un x-\un x_0,\alpha)\,
\Psi^*_\psi\ast\Psi^*_{g^*}(\un y-\un y_0,\alpha)\,
\,\mathcal{M}(\un x, \un x_0\, \un y, \un y_0)\,
e^{-i\un p\cdot (\un z_1-\un z_2)}\nonumber\\
&&
\times
\delta((\un x_0-\un z_1)+\alpha(\un x-\un x_0))\,
\delta((\un y_0-\un z_2)+\alpha(\un y-\un y_0))
\end{eqnarray}

Equation \eq{main_resc2} gives the differential cross section for $\jpsi$ 
production in pA collisions in a quasi-classical approximation to the 
nuclear color field, at large $N_c$ limit and neglecting relativistic 
effects in the $\jpsi$ wave function. The inclusion of high energy 
quantum evolution effects is  important for phenomenological 
applications at RHIC, but is a quite difficult problem. Fortunately, 
the \emph{total} inelastic cross section is not very sensitive to the 
evolution effects as we argue below. Another important reason to 
focus on the total cross section is that it is much less dependent on a 
model which we choose to describe the vector meson wave function.

The total $\jpsi$ production cross section per unit rapidity is found by  
integration over the transverse momentum $\un p$ in \eq{main_resc2}. It 
yields the delta function $(2\pi)^2\delta(\un z_1-\un z_2)$.
It is convenient to introduce the $c\bar c$-dipole transverse 
separation two-vector in the amplitude $\un r=\un x -\un x_0$ and the 
complex-conjugate one $\un r'=\un y-\un y_0$. Now, upon 
substitution of $\mathcal{M}$ into \eq{main_resc2} we get
\begin{eqnarray}
\frac{d\sigma}{dy}&=&
\int d^2b\, \int\,d^2z\, \frac{\as\, C_F}{\pi^2}\,
\frac{1}{\un z^2}\,
\int \frac{d^2r\, d^2 r'}{(2\pi)^2}\,     
\int \frac{d^2x_0\, d^2 y_0}{(2\pi)^2}
\int d\alpha
\nonumber\\
&&
\times\,\Psi_\psi\ast\Psi_{g^*}(\un r,\alpha)\,
\Psi^*_\psi\ast\Psi^*_{g^*}(\un r',\alpha)\,
\bigg(1-
e^{-\frac{1}{4}\big(\frac{Q_s}{2}\big)^4 [(\un x_0-\un y_0)^2+
(\un r-\un r'+\un x_0-\un y_0)^2]^2}
\bigg)\nonumber\\
&&
\times\,\delta(\un x_0-\un z+\alpha \,\un r)\,
\delta(\un y_0-\un z+\,\alpha\, \un r')\label{tot_resc1}\\
&=& \int d^2b\, \int\,d^2z\, \frac{\as\, C_F}{\pi^2}\,
\frac{1}{\un z^2}\,
\int \frac{d^2r\, d^2 r'}{(2\pi)^4}\,
\int d\alpha\,
\Psi_\psi\ast\Psi_{g^*}(\un r,\alpha)\,
\Psi^*_\psi\ast\Psi^*_{g^*}(\un r',\alpha)\,
\nonumber\\
&&
\times\,
\bigg( 1-
e^{-\frac{1}{4}\big(\frac{Q_s}{2}\big)^4 [(\alpha^2+(1-\alpha)^2)(\un 
r-\un r')^2]^2}
\bigg)\label{tot_resc2}
\end{eqnarray}
Using an explicit formula for the overlap function \eq{lcg} in 
\eq{tot_resc2} we evaluate $\un r'$ integral in the approximation $\un 
r'\ll \un r$. The result is 
\beq\label{xcR}
\frac{d\sigma}{dy}\,=\,
S_A\, xG(x_1,Q^2)\,\frac{3\,\Gamma_{ee}}{(2\pi)^2\, 
48\,\alpha_\mathrm{em}\,
M_\psi}\,\int_0^\infty d\zeta\, \zeta^5\, K_2(\zeta)\,
\bigg(1-e^{-\big(\frac{Q_s(x_2)\,\zeta}{2\, M_\psi}\big)^4}\bigg)\,.  
\eeq
where we used \eq{xG} and introduced a dimensionless variable $\zeta=m\, 
r$. The gluon distribution in the proton $xG$ is evaluated at the scale 
$Q^2\,=\,M_\psi^2\,+\,\kappa\,A^{1/3}$, with $\kappa$ a parameter to be 
fixed by experimental data.  For simplicity we assumed that the 
nucleus profile function is $T(b)=\theta(R_A-b)$ where $R_A$ is an effective, centrality-dependent radius 
determined by  the Glauber analysis of $pA$ and $dA$ interactions, see e.g. \cite{Kharzeev:2002ei}.

\section{Interplay of two scales: $M_\psi$ and  
$Q_s$.}\label{sec:interplay} 

\subsection{The effect of quantum evolution}

Equation \eq{xcR} gives the desired result for the 
total cross section of $\jpsi$ production in the quasi-classical 
approximation. In this approximation the saturation scale $Q_s$ is given 
by \eq{satscale}. 
As the result of quantum evolution the saturation scale 
acquires its energy/rapidity dependence:
\beq\label{ssenrgy}
Q_s^2(s,y)=\Lambda^2\, 
A^{1/3}\,\left(\frac{s}{\Lambda^2}\right)^{\frac{\lambda}{2}}\, e^{\lambda 
y}\,, 
\eeq
where $y$ is the rapidity measured in the center-of-mass frame. The value of 
$Q_s^2$ thus increases from the initial value $\Lambda^2\, 
A^{1/3}$ given by \eq{satscale}. The value of $\Lambda\,=\,0.63$~GeV is 
fixed by DIS data.  The rate of 
increase is set by the factor $\lambda$. It is constant in the leading 
logarithmic 
approximation $\lambda=4\as\,N_c/\pi$ \cite{LT}. Various inclusive 
quantities at small $x$ at RHIC and HERA are well fitted with 
$\lambda\approx 0.25$ \cite{GBW,KL,KN,Kharzeev:2002ei,KKT,KLM}. This is close to the 
value one obtains from the Renormalization Group improved \cite{Trian} BFKL 
equation \cite{BFKL}. Moreover, the value of  $\lambda$ 
is approximately constant in the relevant for us range of virtualities, so we assume that 
$\lambda(M_\psi)\approx 0.25$.

In our discussion we assume that evolution of the gluon density in the proton is 
linear since the saturation scale in the proton $Q_{sp}^2$ is $A^{1/3}$ 
times smaller 
than in the nucleus. This is a  justified approximation in the RHIC 
kinematical region where the values of $x_1$ 
are such that the proton wave function is dominated by the Fock states with a relatively 
small number of gluons. However, at the LHC the gluons in the proton will also likely be 
saturated. We will discuss the implications of this in Sec.~\ref{sec:hidden}.

\subsection{Enhancement versus suppression of $\jpsi$ production}
\label{sec:evs}

Let us now investigate the $\jpsi$ production cross section in two 
kinematical regions: (i) $M_\psi>Q_s$ and (ii) $M_\psi <Q_s$. The nuclear 
effects are usually expressed 
in terms of the nuclear modification factor defined as
\beq\label{nmf}
R_{pA}(\jpsi)=\frac{d\sigma^{pA}/dy}{A\,d\sigma^{pp}/dy}\,. 
\eeq
The $\jpsi$ production cross section in pp collisions can be 
obtained by expanding the exponent in \eq{xcR} 
\beq\label{pp}
\frac{d\sigma^{pp}}{dy}=
S_p\, xG(x_1,Q^2)\,\frac{240\,\Gamma_{ee}}{(2\pi)^2\,
\alpha_\mathrm{em}\, M_\psi}\, \bigg(\frac{Q_{sp}(y)\,}{ 
M_\psi}\bigg)^4\,,
\eeq
where $Q_{sp}$ is given by \eq{ssenrgy} with $A=1$.

In the region (i) we can expand the exponent in \eq{xcR} and using 
\eq{pp} we find
\beq\label{nmf2}
R_{pA}(\jpsi)\,=\, A^{1/3}\,\sim\, N_\mathrm{coll}^{Au}\,,\quad M_\psi \gg 
Q_s\,,
\eeq
which means that the $\jpsi$ production is \emph{enhanced} at backward 
rapidities at RHIC and it is stronger for central events than for 
peripheral. 

In the region (ii) the exponent in \eq{xcR} can be neglected which yields
\beq\label{pAb}
\frac{d\sigma^{pA}}{dy}\,=\,
S_A\, xG(x_1,Q^2)\,\frac{6\,\Gamma_{ee}}{(2\pi)^2\,
\alpha_\mathrm{em}\, M_\psi}\,.
\eeq
With the help of \eq{pp} we find the behavior of the nuclear modification 
factor at high energies/forward rapidities
\beq\label{nmf1}
R_{pA}(\jpsi)\,=\,\frac{M_\psi^4\, xG(x_1,Q_A^2)}{40\, A^{1/3}\, 
xG(x_1,Q_p^2)\,
Q_{sp}^4(x_2)}\,\sim 
\frac{e^{-2\,\lambda\,y}}{s^\lambda\,N_\mathrm{coll}},\quad M_\psi \ll 
Q_s\,,
\eeq
where we the subscripts $A$ and $p$  are introduced to distinguish $Q$ in nucleus and proton. It gets \emph{suppressed} both as a function of energy/rapidity and 
centrality. 

\subsection{Limiting fragmentation of $\jpsi$ and \emph{hidden parton 
scaling}}\label{sec:hidden}

\subsubsection{Total cross section}

As has been mentioned in the Introduction a naive collinear factorization
approach implies that the total $\jpsi$ production cross section (as well as the cross section of any other hard process) 
is proportional to the product of parton distribution functions of the proton
$xf(x_1)$ and of the nucleus $xf_A(x_2)$. Moreover, if the coherent effects
are neglected, then $xf_A(x_2)\,=\,A\,xf(x_2)$, i.\ e.\ the nuclear effect
factorizes out. Therefore, in the collinear factorization the total 
cross section is proportional to
\beq\label{coll}
\frac{d\sigma(y)}{dy}\bigg|_\mathrm{coll}\,\propto\,xf(Q\,e^y/\sqrt{s})
\,xf(Q\, e^{-y}/\sqrt{s})\,,
\eeq
where $Q$ is a typical scale of the hard process. At some other energy 
$\sqrt{s'}$ the cross section is:
\beq\label{coll2}
\frac{d\sigma(y)}{dy}\bigg|_\mathrm{coll}\,
\propto\,xf(Q\,e^{y\,+\,Y}/\sqrt{s'})\,xf(Q\, 
e^{-y\,+\,Y}/\sqrt{s'})\,,
\eeq
where $Y\,=\,(1/2)\,\ln(s'/s)$.
However, such dependence on energy  contradicts experimental data on 
inclusive particle production in which 
$d\sigma(y)/dy$ exhibits scaling with $Y+y$ in forward region $y>0$. 
Analogous phenomenon for soft processes is known 
as the ``limiting fragmentation". 

The scaling of $d\sigma(y)/dy$ with $Y+y$ or, equivalently, with $x_1$ is a 
natural consequence of saturation of the nuclear wave function at $y>0$. 
It means that the cross section has the same shape as a function of $x_1$ 
at different energies, see \fig{fig:ydepen}(b). 
In the case of $\jpsi$ production it is manifest in Eq.~\eq{pAb}, see 
\fig{fig:ydepen}(b). We can see that wee partons of the nuclear wave function 
are saturated and hence do not contribute to the fragmentation in the 
forward rapidity region. The parton scaling in the nucleus is effectively 
hidden at small $x_2$; we thus use the term \emph{hidden parton scaling} to describe the 
universal scaling in $x_1$. Since $x_F=x_1$ when $x_2\ll 1$ the hidden parton 
scaling of $d\sigma(y)/dy$ is equivalent to $x_F$ scaling in the same 
kinematical region.

\subsubsection{Nuclear modification factor}

As we noted above, the $A$-dependence of the cross section in the collinear 
factorization approach is trivial. Consequently, the nuclear modification 
factor \eq{nmf} equals unity; it thus scales with $x_2$ and $x_1$. 
Coherent scattering of the proton in the nucleus (manifesting itself in the dependence of the 
scale $Q$ on $A$) breaks this scaling.  
This can be seen in Eq.~\eq{nmf1}, which shows an explicit 
dependence on $x_1$ and $x_2$. Note that the difference between 
$xG(x_1,Q_A^2)$ and $xG(x_1,Q_p^2)$ is largest in the proton fragmentation 
region $x_1\rightarrow 1$ corresponding to very low $x_2$. Here we expect the 
strongest violation of $x_2$ scaling in agreement with experimental data 
\cite{RdeCass}. 

If we compare $R_{pA}(\jpsi)$ at close energies we find an approximate  
$x_F$ scaling. 
This scaling originates in the slow dependence of  the gluon distribution on energy, $Q_{sp}^2\sim xG\sim  s^{\lambda/2}$;  as a result, both  
$x_1$ and $x_F$ scalings approximately hold for close energies, see  \fig{fig:xf}. (The scaling is broken at very forward rapidities $x_1\to 1$, where the sensitivity to the variation of the scale $Q$ is enhanced by the power fall-off of the 
parton distributions, $(1-x_1)^n$.) 
 We thus  explain the $x_F$ scaling observed for the SPS and Fermilab fixed target  
energies. Variation of $xG$ between SPS and RHIC energies produces a much stronger violation of $x_F$ scaling as seen in \fig{fig:xf}. 

In our discussion so far we have always assumed that the proton wave function never
saturates.  However, the gluon saturation in the proton has been likely observed at
HERA (see e.\ g.\ Ref.~\cite{HERA}) at $x$'s somewhat smaller than those 
accessible at RHIC. However, at the LHC much smaller values of $x_2$
can be reached at both central and forward rapidities and the proton wave 
function may form the Color Glass Condensate. In that case $R_{pA}$ 
becomes a function of only $x_1$. This statement is general for all hard 
process, the $\jpsi$ production being a particular example. We thus predict that 
the \emph{ hidden parton scaling}, or the scaling in $x_1$, will become a universal feature at 
the LHC energies.

\section{Comparison with RHIC data}\label{sec:compar}

\subsection{A model}\label{sec:model}

We will now compare our calculations with the experimental data using  
Eq.~\eq{xcR} which is the total $\jpsi$ production cross section in the 
quasi-classical approximation. To take into account the quantum evolution 
effects 
on the scattering amplitude we use a model suggested in \cite{KKT}. 
It gives a good description of inclusive particle production at RHIC. In 
that model the quantum effects in the scattering amplitude are  
parametrized as
\beq\label{model}
1\,-\,e^{-\,\Omega^2}\,
\rightarrow\,1\,-\,e^{-\,\Omega^{2\gamma(y,Q^2)}}\,
\eeq
where $\Omega\,=\,\frac{1}{4}\,Q_s^2(x_2)\,r^2$, and $\gamma(y,Q^2)$ is 
the 
anomalous dimension. Its explicit expression can be found in 
Ref.~\cite{KKT}. Let us only note here that $\gamma$ is chosen in such a way 
as to satisfy the large $Q$, fixed $y$ as well as large $y$, fixed $Q$ 
asymptotic of the DGLAP and the BFKL equations up to the NNLO 
terms. 

Let us now list important assumptions which we made in deriving 
\eq{xcR}. First, the scattering amplitude is calculated in the large $N_c$ 
approximation. Second, in our calculations we assumed that the $\jpsi$ 
wave  function is  non-relativistic. In this approximation the $g^*-\jpsi$ 
overlap function 
takes a simple form shown in \eq{lcg}. Relativistic corrections due to Fermi 
motion strongly depend on the charmed quark mass. On other hand in the 
case of diffractive $\jpsi$ production in DIS it was observed that these 
corrections are almost energy independent \cite{FKS}. Hence, we consider 
the effect of Fermi motion as an uncertainty of the wave function 
normalization \cite{GFLMN} $K_F^2$. 

Third, in  derivation of \eq{xcR} we assumed that only $c$ and $\bar c$ 
interact with the nucleus. Of course the valence quark and the gluon can 
interact 
as well. These processes give a contribution of the same order as 
the one we discussed in the previous subsection, and their parametric 
dependence on energy and atomic number is the same. Therefore, we can also take 
them into account in the overall normalization factor $K_F^2$ \cite{GLLMN}.

Finally, there are parametrically small corrections to Eq.~\eq{xcR} 
due to contributions of the real part of the amplitude and 
off-diagonal matrix elements. These corrections are numerically 
insignificant and we have neglected them.

\subsection{Attenuation in a cold nuclear matter}\label{sec:att}

We have argued in Sec.~\ref{sec:vs} that our result \eq{xcR} is applicable at 
rapidities $y\ge 0$ at RHIC. If we would like to compare it with the lower 
energy data we need to include the effect of absorption in the cold 
nuclear matter due to finite coherence length. 
In that case the produced $c\bar c$ pair and later 
$\jpsi$ itself can inelastically interact with the nuclear matter. This 
reduces the cross section \eq{xcR} by a factor (see e.g. \cite{KZ}) 
\beq\label{supr}
S_\psi\,=\,e^{-\,\sigma^{\psi N}\,\rho\,L(b)}\,,
\eeq
where $\sigma^{\psi N}$ is the inelastic $c\bar c$ ($\jpsi$)--nucleon 
cross section, $\rho=0.17$~fm$^{-3}$ is the nuclear density and 
$L(b)$ is the length traversed by 
$\jpsi$  in nuclear matter at a given impact parameter $\un b$.	
The nuclear absorption factor $S_\psi$ has been studied in detail in the framework of Glauber approach 
Ref.~\cite{Kharzeev:1996yx}. We normalize $S_\psi$ in 
\eq{supr} accordingly.

\subsection{Results of numerical calculations}

We have performed a numerical calculation using the model described in 
Sec.~\ref{sec:model}. The parameter $\kappa$ has been fixed
at $\kappa=0.2$~GeV$^2$, consistent with the analysis of \cite{KN,KL}.
The overall normalization factor $K_F$ in the production cross section has been fitted 
to the RHIC data from  \cite{Adler:2005ph,RdeCass}.

\begin{figure}
\begin{tabular}{ccc}   
\epsfig{file=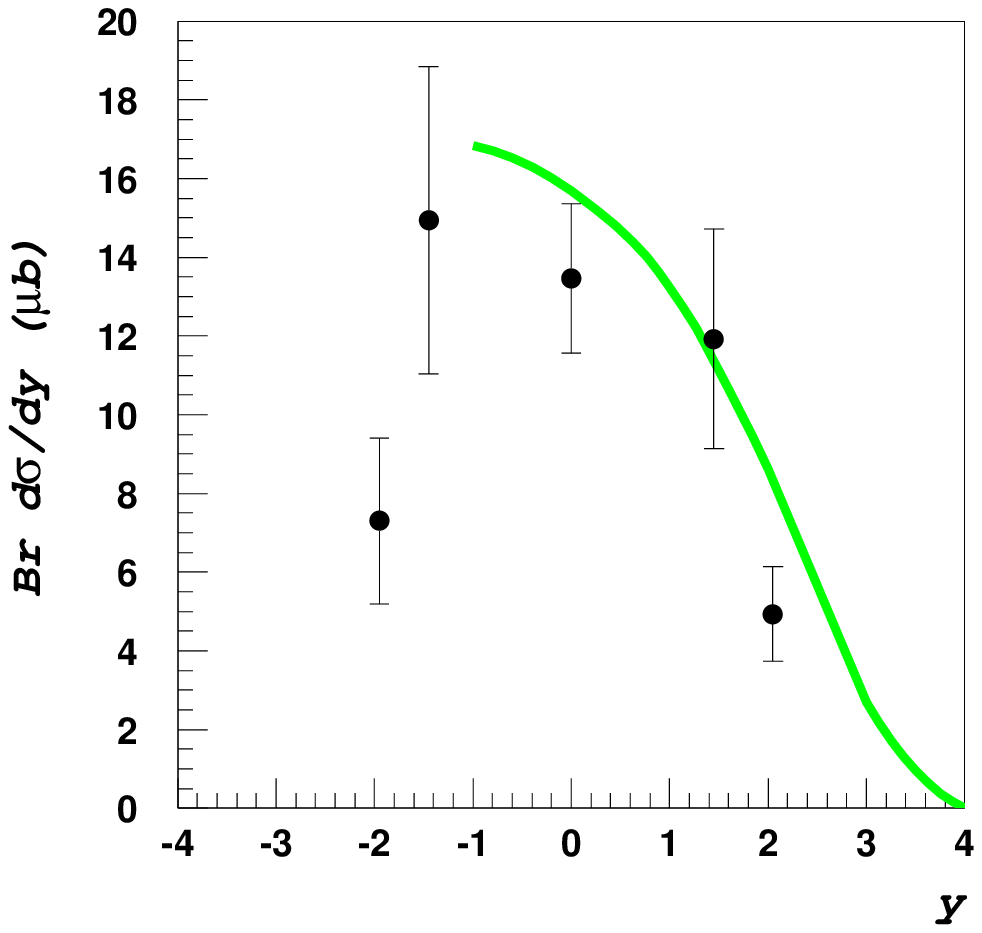,height=9cm}& \mbox{}\hspace{1cm}&
\epsfig{file=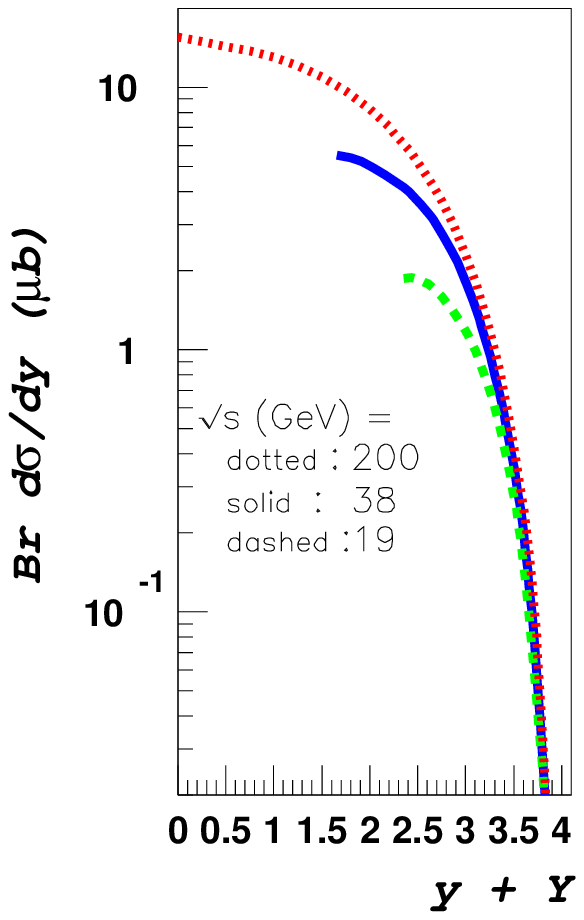,height=9.5cm}\\
(a)&&(b)
\end{tabular}
\caption{ Total inelastic cross section for $\jpsi$ production \eq{xcR} 
as  a function of (a) rapidity $y$, (b) shifted rapidity $y+Y$, where 
$Y=(1/2)\ln(s'/s)$. Data points
are taken from Ref.~\cite{RdeCass}. }
\label{fig:ydepen}
\end{figure}

The result of numerical calculations of the total $\jpsi$ cross section as a
function of rapidity is shown in \fig{fig:ydepen}(a). We observe a reasonable 
agreement with the PHENIX preliminary data \cite{Adler:2005ph,RdeCass}. 

In \fig{fig:ydepen}(b) we show the ``hidden parton scaling" due to the 
saturation of the nuclear wave function, see Sec.~\ref{sec:hidden}. We 
observe an effect similar to the ``limiting fragmentation" in the total 
inclusive cross section. Let us emphasize that such a ``hidden parton 
scaling" is a general feature of all hard processes at high energy in the saturation picture. It 
is due to the saturation of partons in the nucleus. 

\begin{figure}
\begin{tabular}{cc}
\epsfig{file=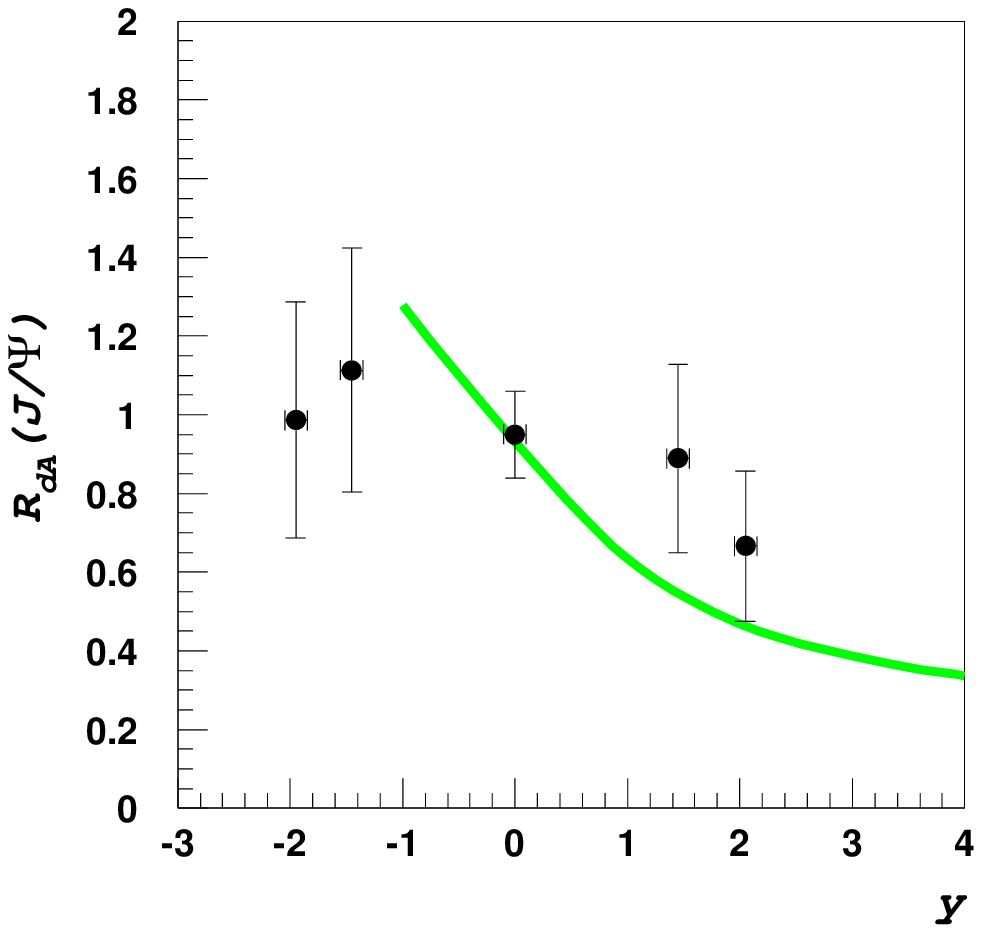,width=8.cm}&
\epsfig{file=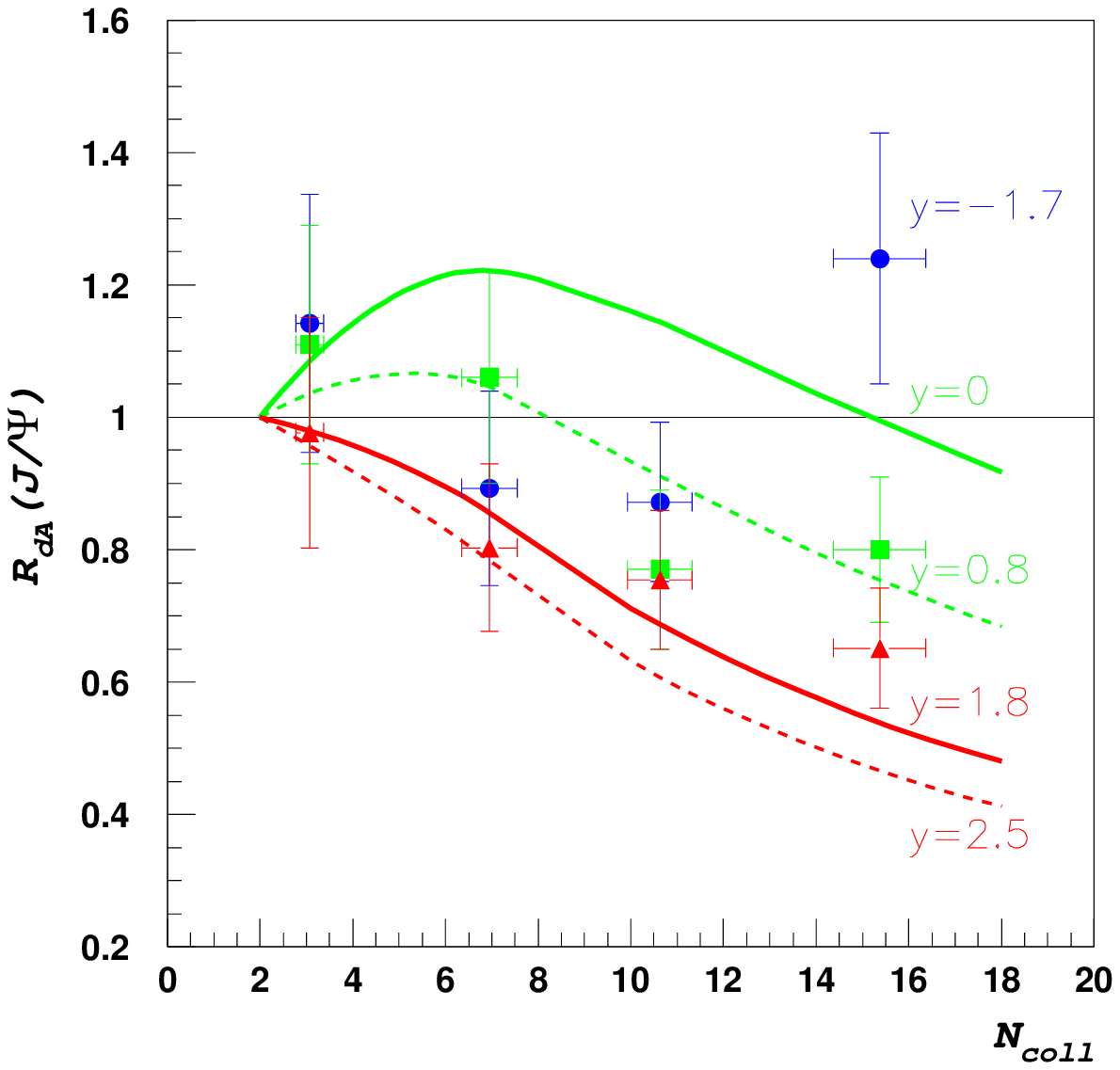,width=8.cm}
\end{tabular}
\caption{Nuclear modification factor $R_\jpsi$ as
a function of (a) rapidity $y$, (b) $N_\mathrm{coll}^{Au}$ at different 
rapidities. Solid lines: numerical calculation. Data points 
are from Ref.~\cite{RdeCass}.}
\label{fig:rdaA}                                                            
\end{figure}

In \fig{fig:rdaA} we show the nuclear modification factor as a function of 
rapidity and centrality. In agreement with our discussion in 
Sec.~\ref{sec:interplay}  $R_{pA}(\jpsi)$ is suppressed at forward 
rapidities and enhanced at midrapidity. The functional dependence on 
rapidity and centrality is given by Eqs.~\eq{nmf2},\eq{nmf1}. Our results 
are in a qualitative agreement with the PHENIX data 
\cite{Adler:2005ph,RdeCass}.

\begin{figure}
\epsfig{file=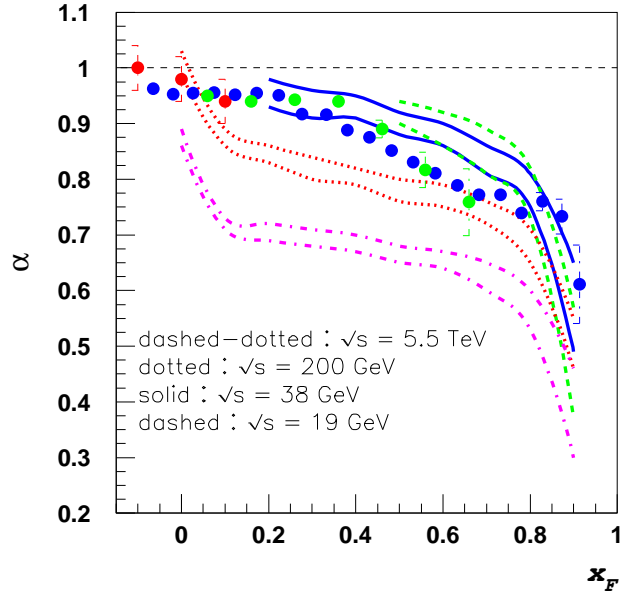,width=10cm}
\caption{$\alpha$ as a function of $x_F$ for different energies. 
$\alpha$ was extracted by fitting function $A^\alpha$ to the calculated 
$\sigma_{pA}/\sigma_{pp}$. Bands show the uncertainty of the fit.}
\label{fig:xf}                                                            
\end{figure}

Finally, in \fig{fig:xf} we present our calculation of the exponent $\alpha$ 
defined as follows: $\sigma_{pA}=A^\alpha\,\sigma_{pp}$. As 
discussed in Sec.~\ref{sec:hidden} we expect an approximate $x_F$ scaling 
at close energies due to the slow dependence of the 
saturation scale on energy. At smaller $x_F$ nuclear absorption plays a  
significant role; because of formation time effects, it suppresses the  $\jpsi$ 
production at SPS stronger 
than at Fermilab and contributes towards improving the $x_F$ scaling.  In the 
same figure we plot our prediction for LHC. At such high energies the 
proton's wave function saturates, which results in the exact $x_1$ scaling, as 
discussed in Sec.~\ref{sec:hidden}. Thus, in the region $x_1\gg x_2$ we 
predict $x_F$ scaling of $\alpha$ at energies $\gsim 5$~TeV. 

We would like to draw the reader's attention to the fact that 
$\alpha\approx 2/3$ at the highest energy in \fig{fig:xf}. This is just the 
value of $\alpha$ one expects to measure in the collision of two black disks: one of 
radius $R_p$ 
another of radius $R_A$. At LHC energies the absorption starts only at $y\le 
-3$. Therefore, from our discussion in Sec.~\ref{sec:evs} we expect that 
$\alpha$  will approach $4/3$ at $y<0$. This behavior will be best seen in 
a plot of $\alpha$ 
as a function of $y$ since the plot versus $x_F$ exponentially expands the 
fragmentation region, while shrinking the interesting central rapidity 
one.

\section{Summary and Conclusions}\label{sec:summary}

In this paper we have analyzed the $\jpsi$ production in p(d)A collisions at high energies. 
We have 
pointed out that due to the large coherence length associated with the 
$\jpsi$ production and formation the dominant production mechanism in the case of heavy nuclei is a 
double-gluon exchange which leads to a significant enhancement of the 
cross section at backward rapidities. In the case of 
DIS this effect has been discussed in Ref.~\cite{HKZ}, and the relevant arguments 
for hard processes at forward rapidities have been given in \cite{Brodsky:1991dj}.   At forward 
rapidities the $\jpsi$ production is suppressed in much the same way as 
open charm is \cite{KhT}. This is  due to saturation of gluons in the nuclear wave 
function, 
see \fig{fig:rdaA}(b). The transition between the enhancement and the suppression regimes in  
$\jpsi$ production  happens when $Q_s\approx M_\psi$. Eq.~\eq{xcR} 
provides an analytical formula for the total cross section of $\jpsi$ 
production in the region where $\tau_P>R_A$.

The dominance of the double-gluon exchange mechanism for the case of heavy nuclei has an important
implication. It means that $\jpsi$ is produced directly in the color singlet state. A clean way to measure 
the contribution of the double gluon exchange from different nucleons  
is to measure $\jpsi$ production in  
proton--deuteron collisions and to trigger on the final fragmentation state of the  
deuteron. A signature of the double-gluon exchange mechanism will be the 
absence of intact nucleons \cite{YD}.  Such an experimental study can shed further 
light on the problem of $\jpsi$ production.  

We have argued that the total $\jpsi$ production cross section exhibits the   
``hidden parton scaling",  see \fig{fig:ydepen}(b). Owing to the saturation of 
 gluons in the nucleus, the cross section becomes independent of $x_2$. Thus it scales 
with $x_1$, or, equivalently, with $y+Y=\ln x_1 + \mathrm{const.}$ In 
other words, $d\sigma(x_1)/dx_1$ has a universal shape for different 
energies.

In Sec.~\ref{sec:hidden} we studied the $x_F$ scaling phenomenon observed 
at lower energies. We found that this scaling holds only 
for close energies since the scaling violating factor is a slow function of 
energy: $\sim s^{0.25}$. However, at energies as high as LHC energy we 
expect 
the saturation of gluons in the proton which will manifests itself in the effective  
disappearance of the factor $s^{0.25}$ and will result in the exact $x_1$ 
scaling. In turn, an exact $x_1$ scaling can be considered as a signature of the  
saturation in the proton.


\acknowledgments

The authors would like to thank Yuri Dokshitzer, Eugene Levin, Larry McLerran,  
Al Mueller and Helmut Satz for informative and helpful discussions.
This research was supported by the U.S. Department of
Energy under Grant No. DE-AC02-98CH10886.
K.T. would like to thank RIKEN, BNL and the U.S. Department of Energy (Contract No. DE-AC02-98CH10886) for providing the facilities essential for the completion of this work.



\begin{thebibliography}{99}


\bibitem{collcharm}
P.~Nason, S.~Dawson and R.~K.~Ellis,
Nucl.\ Phys.\ B {\bf 327}, 49 (1989)
[Erratum-ibid.\ B {\bf 335}, 260 (1990)];
P.~Nason, S.~Dawson and R.~K.~Ellis,
Nucl.\ Phys.\ B {\bf 303}, 607 (1988).

\bibitem{Appelquist:tg}
T.~Appelquist and J.~Carazzone,
Phys.\ Rev.\ D {\bf 11}, 2856 (1975).

\bibitem{Brambilla:2004wf}
  N.~Brambilla {\it et al.},
  arXiv:hep-ph/0412158.

\bibitem{Matsui:1986dk}
T.~Matsui and H.~Satz,
Phys.\ Lett.\ B {\bf 178}, 416 (1986).

\bibitem{Badier:1983dg}
  J.~Badier {\it et al.}  [NA3 Collaboration],
  Z.\ Phys.\ C {\bf 20}, 101 (1983).

\bibitem{Leitch:1999ea}
  M.~J.~Leitch {\it et al.}  [FNAL E866/NuSea collaboration],
  Phys.\ Rev.\ Lett.\  {\bf 84}, 3256 (2000)
  [arXiv:nucl-ex/9909007].

\bibitem{Adler:2005ph}
  S.~S.~Adler {\it et al.}  [PHENIX Collaboration],
  arXiv:nucl-ex/0507032.

\bibitem{Brodsky:1991dj}
  S.~J.~Brodsky, P.~Hoyer, A.~H.~Mueller and W.~K.~Tang,
  Nucl.\ Phys.\ B {\bf 369}, 519 (1992).

\bibitem{Kopeliovich:1984bf}
  B.~Z.~Kopeliovich and F.~Niedermayer,
JINR-E2-84-834
  
\bibitem{Vogt:1991qd}
  R.~Vogt, S.~J.~Brodsky and P.~Hoyer,
  Nucl.\ Phys.\ B {\bf 360}, 67 (1991).

\bibitem{Gavin:1991qk}
  S.~Gavin and J.~Milana,
  Phys.\ Rev.\ Lett.\  {\bf 68}, 1834 (1992).

\bibitem{Kharzeev:1993qd}
  D.~Kharzeev and H.~Satz,
  Z.\ Phys.\ C {\bf 60}, 389 (1993).

\bibitem{Clavelli:1985kg}
  L.~Clavelli, P.~H.~Cox, B.~Harms and S.~Jones,
  Phys.\ Rev.\ D {\bf 32}, 612 (1985).

\bibitem{Benesh:1994du}
  C.~J.~Benesh, J.~w.~Qiu and J.~P.~Vary,
  Phys.\ Rev.\ C {\bf 50}, 1015 (1994)
  [arXiv:hep-ph/9403265];
J.~w.~Qiu, J.~P.~Vary and X.~f.~Zhang,
  Phys.\ Rev.\ Lett.\  {\bf 88}, 232301 (2002)
  [arXiv:hep-ph/9809442].
  
\bibitem{Fujii:2003ff}
  H.~Fujii,
  Prog.\ Theor.\ Phys.\ Suppl.\  {\bf 151}, 127 (2003)
  [arXiv:hep-ph/0303219].


\bibitem{GLR}
L.~V.~Gribov, E.~M.~Levin and M.~G.~Ryskin,
Phys.\ Rept.\  {\bf 100}, 1 (1983).


\bibitem{Mueller:wy}
A.~H.~Mueller and J.~w.~Qiu,
Nucl.\ Phys.\ B {\bf 268}, 427 (1986).

\bibitem{Blaizot:nc}
J.~P.~Blaizot and A.~H.~Mueller,
Nucl.\ Phys.\ B {\bf 289}, 847 (1987).

\bibitem{MV}
L.~D.~McLerran and R.~Venugopalan,
Phys.\ Rev.\ D {\bf 49}, 2233 (1994)
[arXiv:hep-ph/9309289],
Phys.\ Rev.\ D {\bf 49}, 3352 (1994)
[arXiv:hep-ph/9311205],
Phys.\ Rev.\ D {\bf 50}, 2225 (1994)
[arXiv:hep-ph/9402335],
Phys.\ Rev.\ D {\bf 59}, 094002 (1999)
[arXiv:hep-ph/9809427].


\bibitem{KhT}
D.~Kharzeev and K.~Tuchin,
Nucl.\ Phys.\ A {\bf 735}, 248 (2004)
[arXiv:hep-ph/0310358].


\bibitem{KTcc}
K.~Tuchin,
arXiv:hep-ph/0401022,
arXiv:hep-ph/0402298.

\bibitem{Gelis:2003vh}
  F.~Gelis and R.~Venugopalan,
  Phys.\ Rev.\ D {\bf 69}, 014019 (2004)
  [arXiv:hep-ph/0310090].


\bibitem{Blaizot:2004wv}
  J.~P.~Blaizot, F.~Gelis and R.~Venugopalan,
  Nucl.\ Phys.\ A {\bf 743}, 57 (2004)
  [arXiv:hep-ph/0402257].

\bibitem{Gelis:2004jp}
  F.~Gelis, K.~Kajantie and T.~Lappi,
  Phys.\ Rev.\ C {\bf 71}, 024904 (2005)
  [arXiv:hep-ph/0409058].

\bibitem{Fujii:2005vj}
  H.~Fujii, F.~Gelis and R.~Venugopalan,
  arXiv:hep-ph/0504047.

\bibitem{LRSS}
E.~M.~Levin, M.~G.~Ryskin, Y.~M.~Shabelski and A.~G.~Shuvaev,
Sov.\ J.\ Nucl.\ Phys.\  {\bf 53}, 657 (1991)
[Yad.\ Fiz.\  {\bf 53}, 1059 (1991)].

\bibitem{Ioffe}
V.~N.~Gribov, B.~L.~Ioffe and I.~Y.~Pomeranchuk,
Sov.\ J.\ Nucl.\ Phys.\  {\bf 2}, 549 (1966)
[Yad.\ Fiz.\  {\bf 2}, 768 (1965)];
B.~L.~Ioffe,
Phys.\ Lett.\ B {\bf 30}, 123 (1969).

\bibitem{Kharzeev:1995id}
  D.~Kharzeev and H.~Satz,
  Phys.\ Lett.\ B {\bf 356}, 365 (1995)
  [arXiv:hep-ph/9504397].

\bibitem{MuBr}
S.~J.~Brodsky and A.~H.~Mueller,
Phys.\ Lett.\ B {\bf 206}, 685 (1988).

\bibitem{KZ}
B.~Z.~Kopeliovich and B.~G.~Zakharov,
Phys.\ Rev.\ D {\bf 44}, 3466 (1991).

\bibitem{Kharzeev:1999bh}
  D.~Kharzeev and R.~L.~Thews,
  Phys.\ Rev.\ C {\bf 60}, 041901 (1999)
  [arXiv:nucl-th/9907021].

\bibitem{NZ}
N.~N.~Nikolaev and B.~G.~Zakharov,
Z.\ Phys.\ C {\bf 49}, 607 (1991).

\bibitem{AMdipole}
A.~H.~Mueller,
Nucl.\ Phys.\ B {\bf 415}, 373 (1994).

\bibitem{YuK}
Y.~V.~Kovchegov,
Phys.\ Rev.\ D {\bf 54}, 5463 (1996)
[arXiv:hep-ph/9605446];
Phys.\ Rev.\ D {\bf 55}, 5445 (1997)
[arXiv:hep-ph/9701229].

\bibitem{HKZ}
J.~Hufner, B.~Kopeliovich and A.~B.~Zamolodchikov,
Z.\ Phys.\ A {\bf 357}, 113 (1997)
[arXiv:nucl-th/9607033].

\bibitem{Gribov:1969zy}
  V.~N.~Gribov,
SLAC-TRANS-0102

\bibitem{AlMu}
A.~H.~Mueller,
Nucl.\ Phys.\ B {\bf 335}, 115 (1990);

\bibitem{feyncalc}
J.~Kublbeck, H.~Eck and R.~Mertig,
  Nucl.\ Phys.\ Proc.\ Suppl.\  {\bf 29A}, 204 (1992),
http://www.feyncalc.org/.

\bibitem{Kopel1}
B.~Z.~Kopeliovich, A.~V.~Tarasov and A.~Schafer,
Phys.\ Rev.\ C {\bf 59}, 1609 (1999)
[arXiv:hep-ph/9808378];



\bibitem{BFGMS}
S.~J.~Brodsky, L.~Frankfurt, J.~F.~Gunion, A.~H.~Mueller and M.~Strikman,
Phys.\ Rev.\ D {\bf 50}, 3134 (1994)
[arXiv:hep-ph/9402283].

\bibitem{RRML}
M.~G.~Ryskin, R.~G.~Roberts, A.~D.~Martin and E.~M.~Levin,
Z.\ Phys.\ C {\bf 76}, 231 (1997)
[arXiv:hep-ph/9511228].

\bibitem{GLLMN}
E.~Gotsman, E.~Levin, M.~Lublinsky, U.~Maor and E.~Naftali,
Acta Phys.\ Polon.\ B {\bf 34}, 3255 (2003).

\bibitem{Kharzeev:1996yx}
  D.~Kharzeev, C.~Lourenco, M.~Nardi and H.~Satz,
  Z.\ Phys.\ C {\bf 74}, 307 (1997)
  [arXiv:hep-ph/9612217].

\bibitem{Kharzeev:2002ei}
  D.~Kharzeev, E.~Levin and M.~Nardi,
  Nucl.\ Phys.\ A {\bf 730}, 448 (2004)
  [Erratum-ibid.\ A {\bf 743}, 329 (2004)]
  [arXiv:hep-ph/0212316]; 
  Nucl.\ Phys.\ A {\bf 747}, 609 (2005)
  [arXiv:hep-ph/0408050].

\bibitem{LT}
E.~Levin and K.~Tuchin,
Nucl.\ Phys.\ B {\bf 573}, 833 (2000)
[arXiv:hep-ph/9908317],
Nucl.\ Phys.\ A {\bf 693}, 787 (2001)
[arXiv:hep-ph/0101275].

\bibitem{GBW}
K.~Golec-Biernat and M.~W\"usthoff,
Eur.\ Phys.\ J.\ C {\bf 20}, 313 (2001)
[arXiv:hep-ph/0102093],
Phys.\ Rev.\ D {\bf 60}, 114023 (1999)
[arXiv:hep-ph/9903358],
Phys.\ Rev.\ D {\bf 59}, 014017 (1999)
[arXiv:hep-ph/9807513].


\bibitem{KN}
D.~Kharzeev and M.~Nardi,
Phys.\ Lett.\ B {\bf 507}, 121 (2001)
[arXiv:nucl-th/0012025].

\bibitem{KL}
D.~Kharzeev and E.~Levin,
Phys.\ Lett.\ B {\bf 523}, 79 (2001)
[arXiv:nucl-th/0108006].

\bibitem{KLM}
D.~Kharzeev, E.~Levin and L.~McLerran,
Phys.\ Lett.\ B {\bf 561}, 93 (2003)
[arXiv:hep-ph/0210332].

\bibitem{KKT}
D.~Kharzeev, Y.~V.~Kovchegov and K.~Tuchin,
Phys.\ Rev.\ D {\bf 68}, 094013 (2003)
[arXiv:hep-ph/0307037].

\bibitem{Trian}
D.~N.~Triantafyllopoulos,
Nucl.\ Phys.\ B {\bf 648}, 293 (2003)
[arXiv:hep-ph/0209121].

\bibitem{BFKL}
E.~A.~Kuraev, L.~N.~Lipatov and V.~S.~Fadin,
Sov.\ Phys.\ JETP {\bf 45}, 199 (1977)
[Zh.\ Eksp.\ Teor.\ Fiz.\  {\bf 72}, 377 (1977)].
I.~I.~Balitsky and L.~N.~Lipatov,
Sov.\ J.\ Nucl.\ Phys.\  {\bf 28}, 822 (1978)
[Yad.\ Fiz.\  {\bf 28}, 1597 (1978)].

\bibitem{FKS}
L.~Frankfurt, W.~Koepf and M.~Strikman,
Phys.\ Rev.\ D {\bf 54}, 3194 (1996)
[arXiv:hep-ph/9509311];
L.~Frankfurt, W.~Koepf and M.~Strikman,
Phys.\ Rev.\ D {\bf 57}, 512 (1998)
[arXiv:hep-ph/9702216].

\bibitem{GFLMN}
E.~Gotsman, E.~Ferreira, E.~Levin, U.~Maor and E.~Naftali,
Phys.\ Lett.\ B {\bf 503}, 277 (2001)
[arXiv:hep-ph/0101142].

\bibitem{HERA}
E.~Gotsman, E.~Levin, M.~Lublinsky, U.~Maor, E.~Naftali and K.~Tuchin,
J.\ Phys.\ G {\bf 27}, 2297 (2001)
[arXiv:hep-ph/0010198].

\bibitem{RdeCass}
R.G. de Cassagnac (PHENIX Collaboration), talk at ``Quark Matter" 2004, 
Oakland, California, January 12--17, 2004.

\bibitem{YD} This argument is due to Yuri Dokshitzer. 


\end{thebibliography}
\end{document}